\documentclass[twocolumn,twoside,prd,floatfix,letterpaper]{revtex4}
\usepackage{ifpdf}
\ifpdf 
  \usepackage[pdftex]{graphicx} 
\else 
  \usepackage{graphicx} 
\fi
\usepackage{amsmath}
\usepackage{amssymb}
\usepackage{fancyhdr}
\usepackage{journals}
\usepackage{color}
\usepackage{rotating}
\usepackage[colorlinks,hyperindex]{hyperref}

\def \H {{\ensuremath{\cal H}}}
\def \SM {{\ensuremath{\text{SM}}}}
\def \GR {{\ensuremath{\text{GR}}}}
\def \Newton {{\ensuremath{\text{Newton}}}}
\def \other {{\ensuremath{\text{other}}}}

\def \Einstein {{\ensuremath{\text{Einstein}}}}
\def \Eddington {{\ensuremath{\text{Eddington}}}}
\newcommand{\farcs}{\mbox{\ensuremath{.\!\!^{\prime\prime}}}}

\begin{document}

\title{A Quantitative Measure of Theoretical Scientific Merit}
\author{Bruce Knuteson}
\homepage{http://bruceknuteson.com/}
\email{knuteson@mit.edu}


\begin{abstract}
Program review in the physical sciences may benefit from a framework within which to quantitatively discuss the scientific merit of a proposed theoretical program of research, and to assess the scientific merit of a particular theoretical paper.  This article interprets a previously proposed measure of experimental scientific merit in a manner appropriate for quantifying the scientific merit of completed and proposed theoretical research.  With this interpretation, the resulting figure of merit represents a proposal for a quantitative measure of total scientific merit.
\end{abstract}

\maketitle
\tableofcontents

\section{Motivation}

A quantitative measure of experimental scientific merit was proposed in Ref.~\cite{ExperimentalScientificMerit:Knuteson:2007tb}.  This article shows that the same measure, appropriately interpreted, can be used as a quantitative measure of theoretical scientific merit.  The measure, encompassing both experimental and theoretical scientific merit, is thus a proposed measure of total scientific merit.

In the context of determining which research programs to pursue, review committees often must decide the relative scientific merits of proposed research directions, both experimental and theoretical.  Similar decisions are made at all levels, charting directions for entire fields, for large collaborations, for individual university groups, and for individual scientists, and over timescales ranging from a decadal plan for a field to how an individual scientist chooses to allocate her next hour of research time.  These issues arise in the discussion of research directions in which the result is not yet known.

A related issue is faced by those assessing the scientific merit of a particular theoretical or experimental paper.  This topic is the subject of much innocuous lunchroom conversation, and more seriously in the evaluation of the organizations and individuals responsible for producing the result.  Even in the most quantitative subfields in the physical sciences, the discussions leading to these decisions and evaluations are notably non-quantitative.

If direct technological applications are possible, the relevant figure of merit should be something like number of lives saved, tons of reduced carbon emissions, or monetary profit.  This article does not address science with immediate technological implications.  The figure of merit constructed in this article is designed to assess theoretical developments whose technological implications are sufficiently remote to be highly uncertain, leaving their primary short term benefit to be scientific rather than technological.

To make this article self-contained, Section~\ref{sec:ExperimentalScientificMeritReview} briefly reviews the quantitative measure of experimental scientific merit previously proposed in Ref.~\cite{ExperimentalScientificMerit:Knuteson:2007tb}.  Section~\ref{sec:TheoreticalScientificMerit} presents an interpretation in which the measure proposed in Ref.~\cite{ExperimentalScientificMerit:Knuteson:2007tb} can also be used to quantify theoretical scientific merit.  Section~\ref{sec:Examples} provides examples showing how this figure of merit can be applied.  Section~\ref{sec:Discussion} discusses potential advantages of adopting this figure of merit in practice.  Section~\ref{sec:Summary} summarizes.

\section{Review of experimental scientific merit}
\label{sec:ExperimentalScientificMeritReview}

The essential idea of Ref.~\cite{ExperimentalScientificMerit:Knuteson:2007tb} is that the value of a particular experimental result in an academic field should be measured by how much is learned from the result.  Equivalently, the value of a result is how surprised you are that the particular result has been obtained.  An experiment confirming an effect already predicted with high confidence does not teach us much, while an experiment producing an unanticipated result can teach us a great deal.  Section II of Ref.~\cite{ExperimentalScientificMerit:Knuteson:2007tb} develops this basic idea into a quantitative measure of experimental scientific merit, borrowing elementary concepts from information theory.  

Adopting the notation of Ref.~\cite{ExperimentalScientificMerit:Knuteson:2007tb}, let $Y=\{y_1,\ldots,y_j,\ldots,y_m\}$ denote a set of qualitatively distinct and mutually incompatible states of knowledge, the $j^{\text{th}}$ of which is generally accepted to be correct with probability $p(y_j)$, and let $X=\{x_1,\ldots,x_i,\ldots,x_n\}$ denote possible outcomes of an experiment, the $i^{\text{th}}$ of which is expected to be realized with probability $p(x_i)$~\footnote{As a specific example in the field of particle physics, consider $Y=\{y_1,y_2\}$, where $y_1$ denotes the state of knowledge in which a Standard Model Higgs boson exists and $y_2$ denotes the state of knowledge in which a Standard Model Higgs boson is known not to exist, and $X=\{x_1,x_2\}$, where $x_1$ denotes a Higgs boson being discovered at the Fermilab Tevatron and $x_2$ denotes no Higgs boson being discovered at the Fermilab Tevatron.  Reference~\cite{ExperimentalScientificMerit:Knuteson:2007tb} suggests $p(y_1)=95\%$, $p(x_1|y_1)=10\%$, and $p(x_1|y_2)=0$.}.  Since the process of normal science relies to a significant degree on a scientific community's shared view of important problems and possible solutions, in practice there is little difficulty defining $X$ and $Y$~\cite{Kuhn}.  The sets $X$ and $Y$ are assumed to be complete and their individual elements orthogonal, so that $p(x_{i_1},x_{i_2}) = p(y_{j_1},y_{j_2}) = 0$ and $\sum_i{p(x_i)} = \sum_j{p(y_j)}=1$~\footnote{Standard Bayesian notation is used.  The joint probability $p(x_i,y_j)$ is a number between zero and unity quantifying the degree of belief that the outcome of the experiment will be $x_i$ and that the true hypothesis is $y_j$.  The conditional probability $p(y_j|x_i)$ is a number between zero and unity quantifying the degree of belief that the true hypothesis is $y_j$, given that the outcome of the experiment is $x_i$.}.  

The evidence the result $x_i$ provides in favor of the state of knowledge $y_j$ is
\begin{equation}
\text{evidence}(y_j|x_i) = \log_{10}{\frac{p(y_j|x_i)}{p(y_j)}}.
\label{eqn:evidence}
\end{equation}
Since in practice the state of knowledge $y_j$ is not known to be correct, the scientific merit of obtaining the experimental result $x_i$ is the evidence the result $x_i$ provides in favor of $y_j$, averaged over possibly correct states of knowledge $y_j$, weighted according to the current expectation that each $y_j$ is correct.  The scientific merit of the experimental result $x_i$ is thus the information gain, where
\begin{eqnarray}
\!\!\!\!\!\text{information gain }(x_i) & \!\!= & \!\!\sum_j{p(y_j|\text{now}) \, \text{evidence}(y_j|x_i)} \nonumber \\
                      & \!\!= & \!\!\sum_j{p(y_j|\text{now}) \, \log_{10}{\frac{p(y_j|x_i)}{p(y_j)}}},
\label{eqn:InformationGain}
\end{eqnarray}
denoting by $p(y_j|\text{now})$ the current expectation that $y_j$ is correct.  Immediately after the result $x_i$ is obtained, $p(y_j|\text{now}) = p(y_j|x_i)$.  The scientific merit of an experiment not yet performed is the expected value of the scientific merit of its potential results,
\begin{eqnarray}
\Delta H & = & \sum_i{p(x_i) \, \text{information gain }(x_i)} \nonumber \\
         & = & \sum_{i,j}{p(y_j,x_i) \, \log_{10}{\frac{p(y_j|x_i)}{p(y_j)}}},
\label{eqn:DeltaInformationEntropy}
\end{eqnarray}
where $\Delta H$ denotes the expected decrease in information entropy associated with the states of knowledge $Y$ upon performing the experiment $X$.

Further derivation and discussion are provided in Section~II of Ref.~\cite{ExperimentalScientificMerit:Knuteson:2007tb}.

\section{Theoretical scientific merit}
\label{sec:TheoreticalScientificMerit}

The experimental figure of merit assumes a complete set of possible states of
knowledge $Y$.  An experimental outcome $x_i$ adjusts the belief $p(y_j)$ that
each state of knowledge $y_j$ is correct.

Similarly, the theoretical figure of merit proposed in this article assumes a
complete set of possible states of knowledge $Y$.  A theoretical result $x_i$
adjusts the belief $p(y_j)$ that each state of knowledge $y_j$ is correct.  The
theoretical figure of merit proposed in this article is thus the same as the
experimental figure of merit proposed in Ref.~\cite{ExperimentalScientificMerit:Knuteson:2007tb}, 
with appropriately interpreted $x_i$.

It is worth emphasizing that in the proposed figure of merit, a theoretical paper does not obtain its value by extending $Y$, which by assumption is a complete set.  A theoretical paper obtains its value by articulating reasons for adjusting beliefs that some subset of states of knowledge are correct, based on simplicity or agreement with existing data.

Any serious proposal for quantifying theoretical scientific merit must satisfy a basic property of self consistency:  the sum of the merits of two separate papers containing a body of experimental and/or theoretical results must equal the merit of a single paper containing the same body of results.  Current popular proxies for scientific merit, such as number of publications and number of citations, violate this basic property of self consistency.  The figure of merit proposed in this article uniquely satisfies this desired additive property, which is used as the basis of the derivation of the experimental figure of merit in Ref.~\cite{ExperimentalScientificMerit:Knuteson:2007tb}.  This additive property allows the figure of merit to be divided by
cost to obtain a well-defined ``bang per buck,'' quantifying scientific value per
dollar of funding.  The bang per buck can in turn be used in the research
portfolio allocation problem faced at all levels in the scientific enterprise.

With the value of both experimental and theoretical results arising from how
they change beliefs on the set $Y$, the figure of merit proposed in this
article allows a quantitative comparison of the scientific merit of
experimental results with the scientific merit of theoretical results.

\section{Examples}
\label{sec:Examples}

Three short examples will serve to clarify the proposed figure of merit.  

Let \SM\ denote the particle physics Standard Model, believed to be correct with probability
$p(\SM)=1/2$.  A new theoretical paper $x$ articulates a model \H, pointing out
consistency with existing experimental data and noting certain elegant features.
The model $\H$ comes out of left field; no other person has published along
remotely similar lines.  The prior expectation $p(\H)$ that \H\ is the correct
model, corresponding to your belief if someone were to describe the outline of
the idea to you on the street without the supporting justification provided by
the paper, is taken to be $p(\H)=10^{-10}$.  The somewhat arbitrary choice of
$p(\H)=10^{-10}$, which corresponds roughly to giving every human being's pet
theory equal weight, will not greatly affect the result.  The sum of
the beliefs of all other models in $Y$ is $1-p(\SM)-p(\H)$.

After the paper is absorbed by the field, the model \H\ is believed to be
correct with probability $p(\H|x)=\epsilon$, the Standard Model is
believed to be correct with probability $p(\SM|x)=1/2-(\epsilon-10^{-10})$, and
the sum of the beliefs of all other models in $Y$ is unchanged.  Assuming $1 \gg
\epsilon \gg 10^{-10}$, the information gain resulting from the paper $x$ is
\begin{eqnarray}
\text{information gain}(x) & = & p(\SM|x) \log_{10}\frac{p(\SM|x)}{p(\SM)} +  \nonumber \\
& & p(\H|x) \log_{10}\frac{p(\H|x)}{p(\H)} \nonumber \\
                          & \approx & ((1/2 - \epsilon)(-2\epsilon\log_{10}{e}) +  \nonumber \\
& & \epsilon(10+\log_{10}{\epsilon}) \nonumber \\
			   & \approx & \epsilon (9.6 +\log_{10}{\epsilon}),
\label{eqn:LisaRandallExample}
\end{eqnarray}
where terms of order $\epsilon^2$ and $10^{-10}$ have been dropped~\footnote{The term $\log_{10}{e}$ appears in this example, which mixes analytical and numerical computation.  Typically a choice of units in Eq.~\ref{eqn:evidence} using the natural logarithm is most convenient for purely analytical work, and a choice of units using logarithm base ten is most convenient for purely numerical work.}.  If the new model \H\ is believed to be correct with probability $p(\H|x)=\epsilon \approx 10^{-4}$, then the scientific merit of the paper calculated from Eq.~\ref{eqn:LisaRandallExample} is roughly {\tt 6e-4}~\footnote{The style of scientific notation in these examples (writing {\tt 6e-4} rather than $6\times10^{-4}$) is used to emphasize the exponent.}.  

In this example, the scale of the scientific merit of the paper $x$ articulating the novel hypothesis $\H$ is set by the posterior belief $p(\H|x)=\epsilon$ that $\H$ is actually correct.  A multiplicative constant of order unity (equal to roughly 6 in this case) credits novelty.  The theoretical scientific merit of many proposed models in particle physics can be calculated in this manner, using appropriate values of $\epsilon$.

The proposal that the scientific merit of a paper $x$ articulating a new hypothesis \H\ should be closely related to the posterior belief $p(\H|x)$ that \H\ is actually correct may seem strange to fields accustomed to using number of citations as a proxy for scientific worth.  The proposed measure should however align with intuition after further reflection.  Most scientists agree that simple (elegant) theories are generally better than complicated (ugly) theories, and that theories in agreement with existing data are generally better than those in disagreement with existing data.  Through Bayes' theorem, simplicity and fidelity are exactly the quantities that determine the posterior belief. 
 Alternatively, one can note that in the end the whole point is getting the right answer.  Any reasonable figure of merit must therefore incorporate, preferably in as direct a manner as possible, the current belief that the proposed hypothesis \H\ is in fact the correct description of nature.

Consider as a second example a paper $x$ describing an improvement to calculation within the Standard Model that results in closer agreement with existing experimental results.  Such a paper $x$ may increase the posterior belief in the Standard Model from $p(\SM)=1/2$ to $p(\SM|x)=1/2+\epsilon$ by removing a discrepancy between Standard Model prediction and data.  This increased posterior belief in the Standard Model comes at the expense of alternative models \H\ that had received attention in part due to this discrepancy.  The paper $x$ has resulted in the reduction of the summed belief of this set of alternative models from $p(\H)=\epsilon$ to $p(\H|x)=\epsilon' \ll \epsilon$.  The information gain resulting from the paper $x$ is
\begin{eqnarray}
\text{information gain}(x) & = & p(\SM|x) \log_{10}\frac{p(\SM|x)}{p(\SM)}  + \nonumber \\
                           &   & p(\H|x) \log_{10}\frac{p(\H|x)}{p(\H)} \nonumber \\
                          & \approx & ((1/2 + \epsilon)(2\epsilon \log_{10}{e}) + \nonumber \\
                          &     & \epsilon' \log_{10}\frac{\epsilon'}{\epsilon}\nonumber \\
			   & \approx & \epsilon \log_{10}{e},
\label{eqn:SteveMrennaExample}
\end{eqnarray}
where terms of order $\epsilon'$ and $\epsilon^2$ have been dropped.  If the Standard Model after the paper $x$ is believed to be correct with probability $p(\SM|x)=1/2+\epsilon=51\%$, then the scientific merit of the paper calculated from Eq.~\ref{eqn:SteveMrennaExample} is roughly {\tt 4e-3}.  

In this example, the scale of the scientific merit of the paper $x$ articulating an improvement in calculating within the Standard Model is set by the increased confidence $p(\SM|x)-p(\SM)=\epsilon$ that the Standard Model is actually correct.  The theoretical scientific merit of many calculational improvements within the particle physics Standard Model can be determined in this way, using appropriate values of $\epsilon$.

As a historical example, consider two crucial occurrences in the development of general relativity:  Einstein's theoretical developments up to 1917, and the 1919 expedition led by Eddington that confirmed general relativity's prediction for the deflection of starlight by the sun.  Expectations in this example are inferred from the historical recounting of Ref.~\cite{Pais}; conclusions drawn should be checked for robustness under reasonable variations in these expectations.  At the end of 1905, after the introduction of special relativity but before the series of papers culminating in the Einstein field equations, a person approaching you on the street and outlining the hypothesis that would become general relativity would have been believed with an expectation of $p(\GR|1905) \approx 10^{-10}$.  Newtonian gravity was expected to be correct with expectation of $p(\Newton|1905)\approx99\%$, and with probability $p(\other|1905)\approx1\%-10^{-10}$ any of a number of other possibilities might have been correct.  After Einstein's papers, the scientific community remained skeptical that general relativity was indeed a correct description of nature.  The scientific community's expectation in 1917 that general relativity would prove to be correct was $p(\GR|1917) \approx 2\%$, leaving $p(\Newton|1917)\approx97\%$ and $p(\other|1917)\approx1\%$.  Eddington's 1919 measurement of the bending of starlight around the sun at angles of $1\farcs98\pm0\farcs30$ at Sobral and Crommelin's measurement of $1\farcs61\pm0\farcs30$ at Principe during the same eclipse were found to be in significantly better agreement with general relativity's prediction of $1\farcs74$ than the Newtonian prediction of $0\farcs87$.  In the eyes of the scientific community this experimental measurement significantly raised the expectation that general relativity is the correct description of nature, resulting in $p(\GR|1919) \approx 90\%$, $p(\Newton|1919) \approx 9\%$, and $p(\other|1919)\approx1\%$~\footnote{Proper Bayesian updating of the expectation that general relativity was correct after the Eddington measurement accounts for the possibility of a mistake in the measurement, and results in signficantly less certainty in the correctness of general relativity than if the quoted error bars are accepted at face value.}.  

Giving Einstein full credit for the changes in beliefs between 1905 and 1917, the theoretical scientific merit of Einstein's general relativity papers in 1917 was $\text{information gain}(\Einstein,1917) =$
\begin{eqnarray}
\ &p(\GR|1917) \log_{10}\frac{p(\GR|1917)}{p(\GR|1905)} + & \nonumber \\
&p(\Newton|1917) \log_{10}\frac{p(\Newton|1917)}{p(\Newton|1905)} + & \nonumber \\
&p(\other|1917) \log_{10}\frac{p(\other|1917)}{p(\other|1905)} & \approx 0.2. \nonumber
\end{eqnarray}
Giving Eddington full credit for the changes in beliefs between 1917 and 1919, the experimental scientific merit of Eddington's measurement in 1919 was $\text{information gain}(\Eddington,1919) =$
\begin{eqnarray}
\ & p(\GR|1919) \log_{10}\frac{p(\GR|1919)}{p(\GR|1917)} + & \nonumber \\
& p(\Newton|1919) \log_{10}\frac{p(\Newton|1919)}{p(\Newton|1917)} + & \nonumber \\
& p(\other|1919) \log_{10}\frac{p(\other|1919)}{p(\other|1917)} & \approx 1.4. \nonumber
\label{eqn:GeneralRelativityExample1919Eddington}
\end{eqnarray}
Upon Eddington's measurement in 1919, the theoretical scientific merit of Einstein's contribution increased to $\text{information gain}(\Einstein,1919) =$
\begin{eqnarray}
\ & p(\GR|1919) \log_{10}\frac{p(\GR|1917)}{p(\GR|1905)} + & \nonumber \\
& p(\Newton|1919) \log_{10}\frac{p(\Newton|1917)}{p(\Newton|1905)} + & \nonumber \\
& p(\other|1919) \log_{10}\frac{p(\other|1917)}{p(\other|1905)} & \approx 7.5. \nonumber
\label{eqn:GeneralRelativityExample1919Einstein}
\end{eqnarray}
The total scientific merit of Eddington and Einstein in 1919 was $\text{information gain}(\Eddington,\Einstein,1919) =$
\begin{eqnarray}
\ & p(\GR|1919) \log_{10}\frac{p(\GR|1919)}{p(\GR|1905)} + & \nonumber \\
& p(\Newton|1919) \log_{10}\frac{p(\Newton|1919)}{p(\Newton|1905)} + & \nonumber \\
& p(\other|1919) \log_{10}\frac{p(\other|1919)}{p(\other|1905)} & \approx 8.9, \nonumber
\label{eqn:GeneralRelativityExample1919EddingtonEinstein}
\end{eqnarray}
which is seen to be equal to the sum of Eddington's and Einstein's individual scientific merits in 1919, as required for self consistency.

This historical example raises several interesting points.  Increasing the belief of a new theory from extremely unlikely ($p(\GR|1905)\sim 10^{-10}$) to very likely ($p(\GR|1919)$ of order unity) is worth 10 points~\footnote{In general, raising a hypothesis \H\ from an initial belief of $p(\H)=\epsilon$ to a belief of order unity is worth $-\log_{10}{\epsilon}$ points.}, distributed according to the evidence, either theoretical or experimental, provided by each contributor.  The scientific merit of Einstein's theoretical work increased substantially from 1917 to 1919 due to Eddington's evidence supporting general relativity's prediction for the bending of starlight by the sun, even though Einstein himself arguably did not do much from 1917 to 1919.  The increase in the scientific merit of Einstein's work after Eddington's result is consistent with Einstein's worldwide fame for general relativity coming after, rather than prior to, Eddington's measurement.  Interestingly, the scientific merit of Eddington's measurement is within an order of magnitude of the scientific merit of Einstein's theoretical development, suggesting that Eddington deserves significantly more credit than typically given in popular recountings of the history of general relativity.  Conclusions such as these are robust under reasonable variations in estimated expectations.

Ref.~\cite{ExperimentalScientificMerit:Knuteson:2007tb} provides a number of additional examples, using the same formalism to quantify the scientific merits of specific completed and proposed particle physics experiments.

\section{Discussion}
\label{sec:Discussion}

The figure of merit proposed in this article is sufficiently different from common practice in many fields that a brief discussion of salient features may be helpful.

The figure of merit proposed in this article is well behaved.  It is appropriately additive, in the sense that the scientific merit of a single article containing two separate results is equal to the sum of the scientific merits of two different articles describing the results separately.  Popular alternative proxies for merit, such as number of publications or number of citations, violate this basic property of self consistency.  This additive property allows the proposed measure of theoretical merit to meaningfully be divided by cost to obtain a measure of scientific bang per buck that can be used directly to optimize a scientific portfolio.  

The proposed measure of theoretical scientific merit is manifestly consistent with the measure of experimental scientific merit previously proposed in Ref.~\cite{ExperimentalScientificMerit:Knuteson:2007tb}.  This figure of merit thus constitutes a quantitative measure of total scientific merit, encompassing both experimental and theoretical work.  The figure of merit provides a framework for quantitatively comparing the scientific merit of theoretical and experimental results, for quantitatively comparing the scientific merit of proposed theoretical research with proposed experiments, and for optimizing a scientific portfolio consisting of both experimental and theoretical research.  As a specific example, in most subfields there has been little quantitative analysis into the question of whether an increase in theoretical funding relative to experimental funding (or vice versa) would increase expected information gain.  The figure of merit proposed in this article provides a framework for such a quantitative analysis.

Intelligent, conscious maximization of a particular figure of merit is expected to result in a world line with higher values of that figure of merit than a world line in which decisions are made according to other, possibly less well defined, criteria.  Intelligent, conscious maximization of expected information gain is therefore expected to result in a world line with greater information gain than a world line in which decisions are made according to other, possibly less well defined, criteria. 

Use of information content or information gain to evaluate the scientific merit of theoretical contributions requires the estimation of beliefs that proposed theories are correct, and the reader may object that the problem of quantifying a theoretical paper's scientific merit has simply been reformulated in terms of the estimation of the beliefs that theories are correct.  At worst, this reformulation significantly changes and focuses the discussion.   The fact that there is not a well-developed literature to point to for the justification of these beliefs emphasizes the fact that until now the importance of estimating these has not been properly recognized in assessing the scientific merit of theoretical contributions.  In most cases, the scientific conclusion drawn from this reformulation will be robust against the variation of these beliefs within their justifiable range.  Tables I and II in Ref.~\cite{ExperimentalScientificMerit:Knuteson:2007tb} suggest that experimental scientific merit per incremental unit cost ranges over several orders of magnitude.  A similar range in theoretical scientific merit per incremental unit cost is expected.

Some readers may object to the very idea of constructing an explicit figure of merit to quantify the scientific merit of theoretical contributions.  These readers should bear in mind that this already is done (implicitly, if not explicitly) every time a decision of resource allocation or promotion is made.  In discussions of funding for proposed research directions and recognition of completed research directions, value judgments are made regarding the theoretical scientific merit of proposals and results.  Such value judgments are a necessary part of the scientific process.  The question is therefore not whether the theoretical scientific merit of proposals and results should be determined, but rather how best to determine it.  It is surely in each field's interest for such evaluations to be made in the sharpest, most open, most quantifiable, and scientifically best motivated framework possible.  

Some readers may object to the specific figure of merit advocated in this article.  Scientists whose theoretical research scores higher under more traditional measures than under the measure proposed here may be expected to be among those voicing the strongest objections.  These readers are challenged to find a scientifically better motivated figure of merit.

\section{Summary}
\label{sec:Summary}

This article interprets the unique measure of experimental scientific merit developed in Ref.~\cite{ExperimentalScientificMerit:Knuteson:2007tb} in a manner appropriate for quantifying theoretical scientific merit.  

The choice of a reasonable quantitative figure of merit for assessing the scientific merit of proposed theoretical research programs can inform and focus program review and accompanying decisions of resource allocation in many subfields of the physical sciences.  The related choice of a reasonable figure of merit for assessing the scientific merit of any particular theoretical contribution can inform the evaluation of those organizations and individuals responsible for its production.

This article advocates that the scientific merit of a completed theoretical contribution should be quantified by the extent to which it changes beliefs in the correctness of competing candidate theories.  Change in belief of the correctness of competing candidate theories is an elementary notion in the context of information theory, quantified by information gain, defined in Eq.~\ref{eqn:InformationGain}.

This article advocates that the amount of information a program of research is expected to provide is the appropriate quantity for assessing the scientific merit of any proposed theoretical research direction.  Expected information gain is a well understood concept in information theory, quantified by a change in information entropy $\Delta H$, defined in Eq.~\ref{eqn:DeltaInformationEntropy}.

The measure of theoretical scientific merit advocated here, although developed with particle physics and cosmology foremost in mind, is expected to apply equally to other physical sciences in which results may be far removed from practical technological application.  This figure of merit may provide a useful quantitative framework within which decisions about future resource allocation can be made.

\acknowledgments

The author has benefited from a number of conversations with Bill Ashmanskas, Georgios Choudalakis, Ray Culbertson, Michael Miller, and Steve Mrenna on topics related to the subject of this article.

\bibliography{theoreticalScientificMerit}

\end{document}